\documentclass[12pt]{article}
\pdfoutput=1

\usepackage{amsmath}
\usepackage{amssymb}
\usepackage{epsfig}
\usepackage{epstopdf}
\usepackage{latexsym}
\usepackage{color}
\numberwithin{equation}{section}
\usepackage[
      colorlinks=false,
      linkcolor=darkblue,  
      urlcolor=blue,    
      filecolor=blue,     
      citecolor=red,
linktocpage=true,
      pdfstartview=FitV,
      bookmarksopen=true    
      ]{hyperref}

\begin{document}

\begin{titlepage}

\centerline{\huge \rm } 
\bigskip
\centerline{\huge \rm The non-SUSY $AdS_6$ and $AdS_7$ fixed points}
\bigskip
\centerline{\huge \rm are brane-jet unstable}
\bigskip
\bigskip
\bigskip
\bigskip
\bigskip
\bigskip
\centerline{\rm Minwoo Suh}
\bigskip
\centerline{\it Department of Physics, Kyungpook National University, Daegu 41566, Korea}
\bigskip
\centerline{\tt minwoosuh1@gmail.com} 
\bigskip
\bigskip
\bigskip
\bigskip
\bigskip
\bigskip
\bigskip
\bigskip
\bigskip
\bigskip

\begin{abstract}
\noindent In six- and seven-dimensional gauged supergravity, each scalar potential has one supersymmetric and one non-supersymmetric fixed points. The non-supersymmetric $AdS_7$ fixed point is perturbatively unstable. On the other hand, the non-supersymmetric $AdS_6$ fixed point is known to be perturbatively stable. In this note we examine the newly proposed non-perturbative decay channel, called brane-jet instabilities of the $AdS_6$ and $AdS_7$ vacua. We find that when they are uplifted to massive type IIA and eleven-dimensional supergravity, respectively, the non-supersymmetric $AdS_6$ and $AdS_7$ vacua are both brane-jet unstable, in fond of the weak gravity conjecture.
\end{abstract}

\vskip 5cm

\flushleft {April, 2020}

\end{titlepage}

\tableofcontents

\section{Introduction and conclusions}

The AdS/CFT correspondence, \cite{Maldacena:1997re}, has provided a framework to study quantum field theories in various dimensions with various amount of supersymmetry through their gravitational duals. When it comes to the non-supersymmetric quantum field theories, even though there are several known perturbatively stable, \cite{Breitenlohner:1982bm, Breitenlohner:1982jf, Gibbons:1983aq}, non-supersymmetric $AdS$ vacua, due to the limited control over the non-supersymmetry, not much was able to be investigated.{\footnote{There is a recent search of curious $5d$ non-supersymmetric CFTs in \cite{BenettiGenolini:2019zth}.}} Furthermore, recently, as a stronger version of the weak gravity conjecture, \cite{ArkaniHamed:2006dz}, a conjecture on non-supersymmetric $AdS$ vacua was suggested: there are no stable non-supersymmetric $AdS$ vacua from string and M-theory, \cite{Ooguri:2016pdq}. In support of testing the conjecture, a new non-perturbative decay channel called $brane$-$jet$ instability was proposed by Bena, Pilch and Warner in \cite{Bena:2020xxb}. This examines the force acting on the probe branes and if the force is repulsive, the vacuum is determined to be unstable. In \cite{Bena:2020xxb}, the authors showed the only known perturbatively stable non-supersymmetric $AdS_4$ vacuum, \cite{Warner:1983du, Warner:1983vz}, among the $AdS$ vacua of four-, \cite{deWit:1982bul}, and five-, \cite{Khavaev:1998fb}, dimensional maximal gauged supergravity is, in fact, brane-jet unstable. See also \cite{Guarino:2020jwv} for the brane-jet stability from the D2-brane theories.

There is a closely related channel of non-perturbative instability from instantons. The condition for nucleation of a bubble in Euclidean $AdS$ is analyzed in \cite{Maldacena:1998uz}. It is given by the competition of tension and charge of the particles. Once the bubble is created, it reaches the boundary of $AdS$ in finite Lorentzian time and destabilizes the $AdS$ spacetime.{\footnote{We would like to thank Gabriele Lo Monaco and collaborators of \cite{Apruzzi:2019ecr} for comments on this.}} This idea was recently extended to branes in string theory in \cite{Apruzzi:2019ecr}. See $e.g.$, \cite{Horowitz:2007pr, Gaiotto:2009mv, Narayan:2010em, Ooguri:2017njy} also for studies of instability of $AdS$ and instantons.

The purpose of this note is to examine the brane-jet instability of $AdS$ vacua of six- and seven-dimensional gauged supergravity in \cite{Romans:1985tw} and in \cite{Townsend:1983kk, Mezincescu:1984ta, Pernici:1984xx, Pernici:1984zw}, respectively. In six- and seven-dimensional gauged supergravity, each scalar potential has one supersymmetric and one non-supersymmetric fixed points. 

In seven dimensions, minimal gauged supergravity, \cite{Townsend:1983kk, Mezincescu:1984ta}, is a subsector of maximal gauged supergravity, \cite{Pernici:1984xx, Pernici:1984zw}. As we identify the scalar fields to a single scalar field, the maximal theory reduces to the minimal theory. The scalar potentials of the theories have a pair of supersymmetric and non-supersymmetric fixed points. The non-supersymmetric fixed point is known to be perturbatively stable in the minimal theory, \cite{Mezincescu:1984ta}, but not stable in the maximal theory, \cite{Pernici:1984zw}. Maximal and minimal theories commonly uplift to eleven-dimensional supergravity, \cite{Nastase:1999cb, Nastase:1999kf, Nastase:2000tu} and \cite{Lu:1999bc}, but the minimal theory also uplifts to massive type IIA supergravity, \cite{Passias:2015gya}. We will examine the brane-jet stability of the $AdS_7$ fixed points when they are uplifted to eleven-dimensional supergravity.

In $F(4)$ gauged supergravity in six dimensions, \cite{Romans:1985tw}, there are also a pair of supersymmetric and non-supersymmetric fixed points. The non-supersymmetric $AdS_6$ fixed point is known to be perturbatively stable, \cite{Romans:1985tw}. $F(4)$ gauged supergravity is a consistent truncation of massive type IIA supergravity, \cite{Cvetic:1999un} and also of type IIB supergravity, \cite{Jeong:2013jfc, Hong:2018amk, Malek:2018zcz}. We will examine the brane-jet stability of the $AdS_6$ fixed points when they are uplifted to massive type IIA supergravity.

Indeed we show that when they are uplifted to massive type IIA and eleven-dimensional supergravity, respectively, the non-supersymmetric $AdS_6$ and $AdS_7$ fixed points are both brane-jet $unstable$ in favor of the conjecture on non-supersymmetric vacua in \cite{Ooguri:2016pdq}.

It would be interesting to consider the alternative uplifts of the $AdS_6$ and $AdS_7$ fixed points to type IIB, \cite{Jeong:2013jfc, Hong:2018amk, Malek:2018zcz} from \cite{DHoker:2016ujz, DHoker:2016ysh}, and massive type IIA supergravity, \cite{Passias:2015gya} from \cite{Apruzzi:2013yva}, respectively. Indeed, the instabilities of $AdS_7$ solutions in massive type IIA supergravity are already examined in \cite{Apruzzi:2016rny, Danielsson:2017max, Apruzzi:2019ecr}.{\footnote{Some massless solutions considered in \cite{Apruzzi:2019ecr} would be obtained from dimensional reduction of the $AdS_7$ solutions of eleven-dimensional supergravity we study in this work.}}

In section 2 and 3, we test the brane-jet instabilities of $AdS$ fixed points from six- and seven-dimensional gauged supergravity, respectively. In an appendix, we present the calculation of potentials of the fluxes for supersymmetric flows and show that the probe brane potentials vanish over the whole flows identically.

\section{The $AdS_6$ fixed points}

\subsection{Solutions  in massive type IIA supergravity}

We consider the scalar-gravity action of $F(4)$ gauged supergravity, \cite{Romans:1985tw}, in the conventions of \cite{Cvetic:1999un},
\begin{equation}
e^{-1}\mathcal{L}\,=\,R-\frac{1}{2}\partial_\mu\phi\partial^\mu\phi-g^2\left(\frac{2}{9}e^{\frac{3}{\sqrt{2}}\phi}-\frac{8}{3}e^{\frac{1}{\sqrt{2}}\phi}-2e^{-\frac{1}{\sqrt{2}}\phi}\right)\,.
\end{equation}
There are supersymmetric and non-supersymmetric fixed points of the scalar potential at $e^{-\frac{1}{2\sqrt{2}}\phi}=1$ and $e^{-\frac{1}{2\sqrt{2}}\phi}=1/3^{1/4}$, respectively.

We consider the domain wall background,
\begin{equation}
ds_6^2\,=\,e^{2A}ds_{1,4}^2+dr^2\,,
\end{equation}
where $A\,=\,r/l$ at the $AdS_6$ fixed points. The radius of $AdS_6$ is given by $l$.

We employ the uplift formula to massive type IIA supergravity, \cite{Romans:1985tz}, in \cite{Cvetic:1999un}. In Einstein frame, the metric, the dilaton, and the four-form flux are non-trivial and are given, respectively, by, \cite{Brandhuber:1999np},
\begin{equation}
ds^2\,=\,X^{1/8}\sin^{1/12}\xi\left(\Delta^{3/8}ds_6^2+\frac{2}{g^2}\Delta^{3/8}X^2d\xi^2+\frac{1}{2g^2}\frac{\cos^2\xi}{\Delta^{5/8}X}ds_{S^3}^2\right)\,,
\end{equation}
\begin{equation}
e^{\Phi}\,=\,\frac{\Delta^{1/4}}{X^{5/4}\sin^{5/6}\xi}\,,
\end{equation}
\begin{equation}
F_{(4)}\,=\,-\frac{\sqrt{2}}{6}\frac{U\sin^{1/3}\xi\cos^3\xi}{g^3\Delta^2}d\xi\wedge{vol}_{S^3}-\sqrt{2}\frac{\sin^{4/3}\xi\cos^4\xi}{g^3\Delta^2X^3}dX\wedge{vol}_{S^3}\,,
\end{equation}
where we define
\begin{align}
X\,=&\,e^{-\frac{\phi}{2\sqrt{2}}}\,, \notag \\
\Delta\,=&\,X\cos^2\xi+X^{-3}\sin^2\xi\,, \notag \\
U\,=&\,X^{-6}\sin^2\xi-3X^2\cos^2\xi+4X^{-2}\cos^2\xi-6X^{-2}\,.
\end{align}
The metric and the volume form of the three-sphere are given, respectively, by
\begin{align}
ds_{S^3}^2\,=&\,\sum^3_{I=1}\left(\sigma^I\right)^2\,, \notag \\
vol_{S^3}\,=&\,\sigma_1\wedge\sigma_2\wedge\sigma_3\,,
\end{align}
where $\sigma^I$ are $SU(2)$ left-invariant one-forms,
\begin{equation}
d\sigma^I\,=\,-\frac{1}{2}\epsilon_{IJK}\sigma^J\wedge\sigma^K\,.
\end{equation}
We may introduce explicit $SU(2)$ left-invariant one-forms,
\begin{align}
\sigma^1\,=&\,-\sin\alpha_2\cos\alpha_3d\alpha_1+\sin\alpha_3d\alpha_2\,, \notag \\
\sigma^2\,=&\,\sin\alpha_2\sin\alpha_3d\alpha_1+\cos\alpha_3d\alpha_2\,,  \notag \\
\sigma^3\,=&\,\cos\alpha_2d\alpha_1+d\alpha_3\,.
\end{align}
Then the metric and the volume form are
\begin{align}
ds_{S^3}^2\,=&\,d\alpha_1^2+d\alpha_2^2+d\alpha_3^2+2\cos\alpha_2d\alpha_1d\alpha_3\,, \notag \\
vol_{S^3}\,=&\,\sin\alpha_2d\alpha_1d\alpha_2d\alpha_3\,.
\end{align}

\subsection{D4-brane probes}

The uplift formula for the six-form flux is given by, \cite{Cvetic:1999un},
\begin{equation}
F_{(6)}\,=\,e^{\Phi/2}*F_{(4)}\,=\,-\frac{\sqrt{2}g}{3}Uvol_6+\frac{4\sqrt{2}}{g}\frac{\sin\xi\cos\xi}{X}*dX\wedge{d}\xi\,+\cdots.
\end{equation}
At the $AdS_6$ fixed points, it gives
\begin{equation}
F_{(6)}\,=\,-\frac{\sqrt{2}g}{3}Ue^{5A}dx_0\wedge{d}x_1\wedge{d}x_2\wedge{d}x_3\wedge{d}x_4\wedge{d}r\,,
\end{equation}
where
\begin{equation}
A\,=\,\frac{r}{l}\,, \qquad U\,=\,U(\xi)\,, \qquad X\,=\,constant\,,
\end{equation}
and $l$ is the radius of $AdS_6$. Thus we obtain that the five-form potential is
\begin{equation}
C_{(5)}\,=\,\frac{\sqrt{2}g}{3}\frac{l}{5}Ue^{5A}dx_0\wedge{d}x_1\wedge{d}x_2\wedge{d}x_3\wedge{d}x_4\,,
\end{equation}
where we use $\partial_rU\,=\,0$, $\partial_\xi{U}\,=\,0$ at the fixed points. $U$ is so-called geometric scalar potential.{\footnote{For the supersymmetric flows we can calculate the five-form potential over the whole flow. See appendix A.1.}}

We partition the spacetime coordinates,
\begin{equation}
x^a\,=\,\{x_0,x_1,x_2,x_3,x_4\}\,, \qquad y^m\,=\,\{r,\xi,\alpha_1,\alpha_2,\alpha_3\}\,,
\end{equation}
and choose the static gauge,
\begin{equation}
x_0\,=\,t\,=\,\eta^0\,,\qquad x^a\,=\,\eta^a\,, \qquad y^m\,=\,y^m(t)\,,
\end{equation}
where $\eta^a$ are the worldvolume coordinates. The pull-back of the metric is
\begin{equation}
\tilde{G}_{ab}\,=\,G_{\mu\nu}\frac{\partial{x}^\mu}{\partial\eta^a}\frac{\partial{x}^\nu}{\partial\eta^b}\,.
\end{equation}

\begin{figure}[t]
\begin{center}
\includegraphics[width=3.0in]{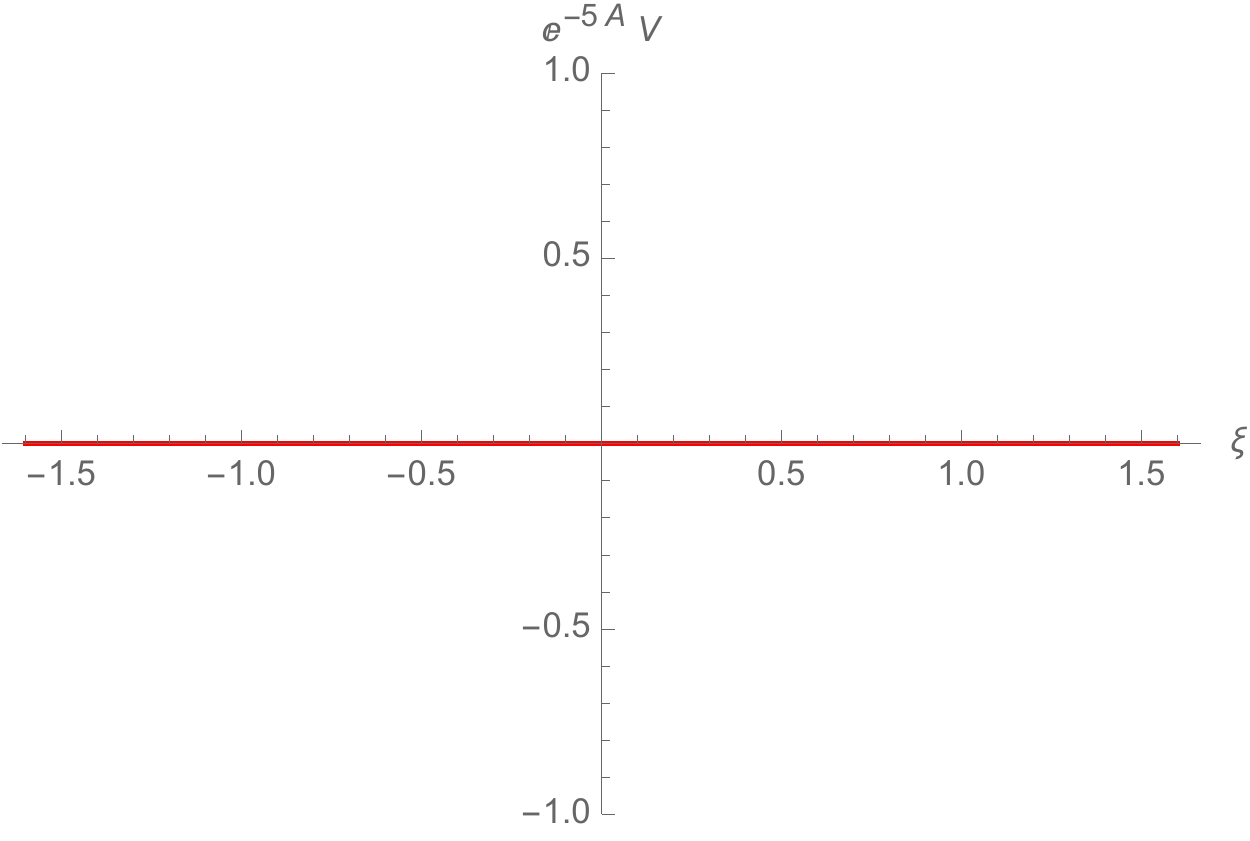} \qquad \includegraphics[width=3.0in]{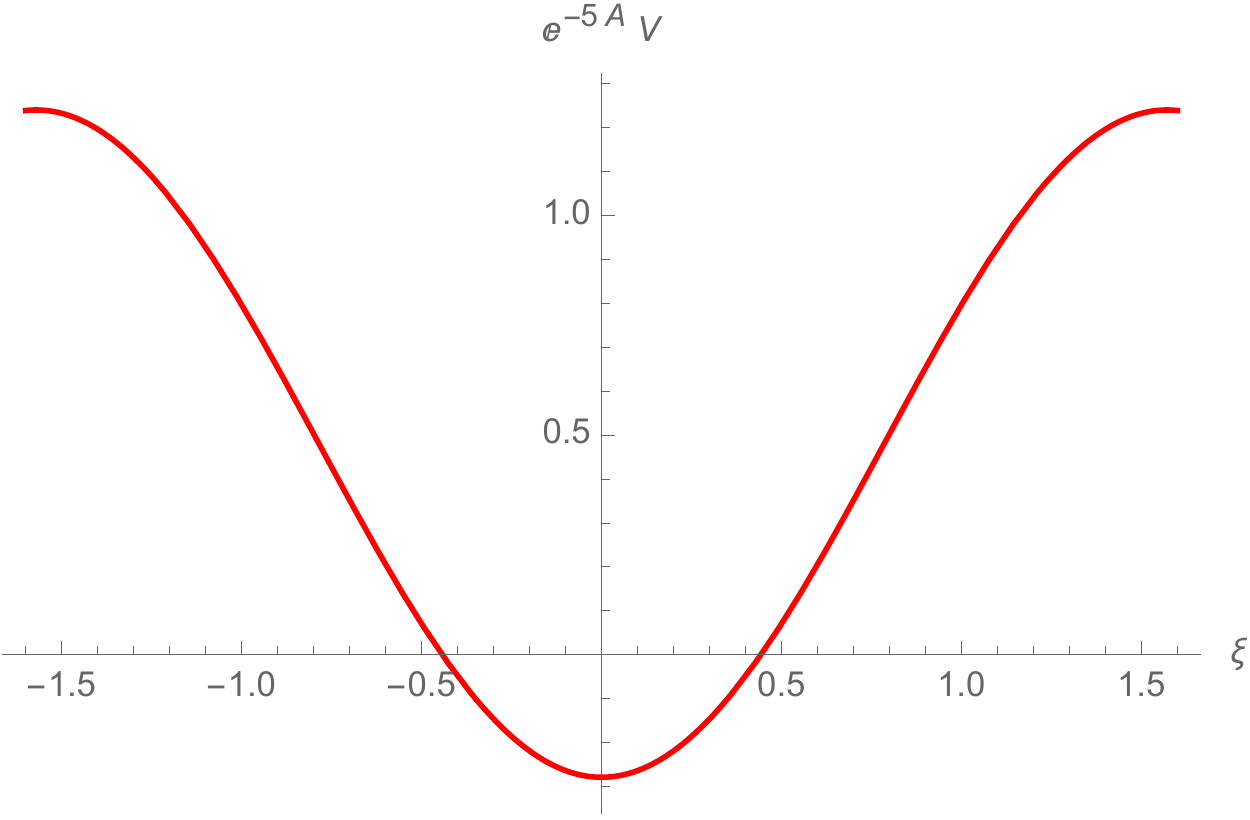}
\caption{{\it The probe brane potentials of the supersymmetric and non-supersymmetric fixed points at $X\,=\,1$ and $X\,=\,1/3^{1/4}$, respectively.}}
\label{1}
\end{center}
\end{figure}

Now we study the worldvolume action of the D4-branes which is given by a sum of DBI and WZ terms. If the probe branes move slowly, the worldvolume action in Einstein frame is
\begin{align}
S\,=&\,-e^{\Phi/4}\int{d}^5\eta\sqrt{-\text{det}(\tilde{G})}-\int\tilde{C}_{(5)} \notag \\
=&\,-\frac{\Delta^{1/16}}{X^{5/16}\sin^{5/24}\xi}\int{d}^5\eta\left(e^{5A}\Delta^{15/16}X^{5/16}\sin^{5/24}\xi-\frac{1}{2}e^{3A}\Delta^{9/16}X^{3/16}\sin^{1/8}\xi{G}_{mn}\dot{y}^m\dot{y}^n+\cdots\right) \notag \\
&\,-\int\frac{\sqrt{2}g}{3}\frac{l}{5}{e}^{5A}U\,{d}x_0\wedge{d}x_1\wedge{d}x_2\wedge{d}x_3\wedge{d}x_4\,,
\end{align}
where $\tilde{C}_{(5)}$ is the pull-back of the five-form potential.{\footnote{The sign of the $\tilde{C}_{(5)}$ term is determined by the orientation of our solution. There is an overall sign choice for the supersymmetry equations in \eqref{susyeq} and it is interelated to the sign choice.}} Then the worldvolume action reduces to
\begin{equation}
S\,=\,\int{d}^5\eta\left(K-V\right)\,,
\end{equation}
where the kinetic and the potential terms are
\begin{align}
K\,=&\,\frac{1}{2}e^{3A}\frac{\Delta^{5/8}}{X^{1/8}\sin^{1/12}\xi}{G}_{mn}\dot{y}^m\dot{y}^n+\cdots\,, \notag \\
V\,=&\,e^{5A}\left(\Delta+\frac{\sqrt{2}g}{3}\frac{l}{5}U\right)\,.
\end{align}
The final probe brane potential is quite simple. From the probe brane potential, we test the brane-jet instabilities of the supersymmetric and non-supersymmetric $AdS_6$ fixed points. We set $g\,=\,\frac{3\sqrt{2}}{2}$ for $l\,=\,1$. The plots of the brane potential over the hemisphere, $0\,\le\,\xi\,\le\,\pi$, are given in Figure 1. We conclude that the non-supersymmetric $AdS_6$ fixed point is $not$ stable.

\section{The $AdS_7$ fixed points}

\subsection{Solutions in eleven-dimensional supergravity}

We consider the minimal scalar-gravity action of seven-dimensional gauged supergravity, \cite{Townsend:1983kk, Mezincescu:1984ta} and \cite{Pernici:1984xx, Pernici:1984zw}, in the conventions of \cite{Cvetic:1999xp},
\begin{equation}
e^{-1}\mathcal{L}\,=\,R-20\partial_\mu\lambda\partial^\mu\lambda+g^2\left(4X^2+4X^{-3}-\frac{1}{2}X^{-8}\right)\,,
\end{equation}
where $X\,=\,e^{2\lambda}$. There are supersymmetric and non-supersymmetric fixed points of the scalar potential at $X=1$ and $X=1/2^{1/5}$, respectively.

We consider the domain wall background,
\begin{equation}
ds_7^2\,=\,e^{2A}ds_{1,5}^2+dr^2\,,
\end{equation}
where $A\,=\,r/l$ at the $AdS_7$ fixed points. The radius of $AdS_7$ is given by $l$.

We employ the uplift formula to eleven-dimensional supergravity, \cite{Cremmer:1978km}, in \cite{Cvetic:1999xp}. The metric and the seven-form flux are given by,
\begin{equation}
ds^2\,=\,\Delta^{1/3}ds_7^2+\frac{1}{g^2}\Delta^{-2/3}\left(X_0^{-1}d\mu_0^2+\sum_{i=1}^2X_i^{-1}\left(d\mu_i^2+\mu_i^2d\phi_i^2\right)\right)\,,
\end{equation}
\begin{equation}
F_{(7)}\,=\,Uvol_7+\frac{1}{2g}\sum_{\alpha=0}^2X_\alpha^{-1}*_7dX_\alpha\wedge{d}(\mu_\alpha^2)\,,
\end{equation}
where $vol_7$ and $*_7$ are volume form and Hodge dual on $ds_7^2$. We define{\footnote{For the scalar fields in \cite{Cvetic:1999xp}, $X_1\,=\,e^{-\frac{1}{\sqrt{2}}\varphi_1-\frac{1}{\sqrt{10}}\varphi_2}$ and $X_2\,=\,e^{\frac{1}{\sqrt{2}}\varphi_1-\frac{1}{\sqrt{10}}\varphi_2}$, we set $\varphi_1\,=\,0$ and $\varphi_2\,=\,-2\sqrt{10}\lambda$.}
\begin{align}
X\,=&\,X_1\,=\,X_2\,=\,e^{2\lambda}\,, \qquad X_0\,=\,(X_1X_2)^{-2}\,, \notag \\
\Delta\,=&\,\sum_{\alpha=0}^2X_\alpha\mu_\alpha^2\,, \notag \\
U\,=&\,2g\sum_{\alpha=0}^2\left(X_\alpha^2\mu_\alpha^2-\Delta{X}_\alpha\right)+g\Delta{X}_0\,, \notag \\
1\,=&\,\sum_{\alpha=0}^2\mu_\alpha^2\,.
\end{align}
We introduce explicit coordinates,
\begin{equation}
\mu_0\,=\,\cos\alpha\,, \qquad \mu_1\,=\,\sin\alpha\cos\beta\,, \qquad \mu_2\,=\,\sin\alpha\cos\beta\,.
\end{equation}

\subsection{M5-brane probes}

At the $AdS_7$ fixed points, the seven-form flux is
\begin{equation}
F_{(7)}\,=\,Ue^{6A}dx_0\wedge{d}x_1\wedge{d}x_2\wedge{d}x_3\wedge{d}x_4\wedge{d}x_5\wedge{d}r\,,
\end{equation}
where
\begin{equation}
A\,=\,\frac{r}{l}\,, \qquad U\,=\,U(\alpha)\,, \qquad X\,=\,constant\,,
\end{equation}
and $l$ is the radius of $AdS_7$. Thus we obtain that the six-form potential is
\begin{equation}
C_{(6)}\,=\,\frac{l}{6}Ue^{6A}dx_0\wedge{d}x_1\wedge{d}x_2\wedge{d}x_3\wedge{d}x_4\wedge{d}x_5\,,
\end{equation}
where we use $\partial_rU\,=\,0$, $\partial_\alpha{U}\,=\,0$ at the fixed points. $U$ is so-called geometric scalar potential.{\footnote{For the supersymmetric flows we can calculate the six-form potential over the whole flow. See appendix A.2.}}

We partition the spacetime coordinates,
\begin{equation}
x^a\,=\,\{x_0,x_1,x_2,x_3,x_4,x_5\}\,, \qquad y^m\,=\,\{r,\alpha,\beta,\phi_1,\phi_2\}\,,
\end{equation}
and choose the static gauge,
\begin{equation}
x_0\,=\,t\,=\,\eta^0\,,\qquad x^a\,=\,\eta^a\,, \qquad y^m\,=\,y^m(t)\,,
\end{equation}
where $\eta^a$ are the worldvolume coordinates. The pull-back of the metric is
\begin{equation}
\tilde{G}_{ab}\,=\,G_{\mu\nu}\frac{\partial{x}^\mu}{\partial\eta^a}\frac{\partial{x}^\nu}{\partial\eta^b}\,.
\end{equation}

Now we study the worldvolume action of the M5-branes which is given by a sum of DBI and WZ terms. If the probe branes move slowly, the worldvolume action is
\begin{align}
S\,=&\,-\int{d}^6\eta\sqrt{-\text{det}(\tilde{G})}-\int\tilde{C}_{(6)} \notag \\
=&\,-\int{d}^6\eta\left(e^{6A}\Delta-\frac{1}{2}e^{4A}\Delta^{2/3}G_{mn}\dot{y}^m\dot{y}^n+\cdots\right) \notag \\
&\,-\int\frac{l}{6}U{e}^{6A}\,{d}x_0\wedge{d}x_1\wedge{d}x_2\wedge{d}x_3\wedge{d}x_4\wedge{d}x_5\,.
\end{align}
where $\tilde{C}_{(6)}$ is the pull-back of the six-form potential.{\footnote{The sign of the $\tilde{C}_{(6)}$ term is determined by the orientation of our solution. There is an overall sign choice for the supersymmetry equations in \eqref{susyeq7} and it is interelated to the sign choice.}} Then the worldvolume action reduces to
\begin{equation}
S\,=\,\int{d}^6\eta\left(K-V\right)\,,
\end{equation}
where the kinetic and the potential terms are
\begin{align}
K\,=&\,\frac{1}{2}e^{4A}\Delta^{2/3}G_{mn}\dot{y}^m\dot{y}^n+\cdots\,, \notag \\
V\,=&\,e^{6A}\left(\Delta+\frac{l}{6}U\right)\,.
\end{align}
The final probe brane potential is quite simple. From the probe brane potential, we test the brane-jet instabilities of the supersymmetric and non-supersymmetric $AdS_7$ fixed points. We set $g\,=\,2$ for $l\,=\,1$. The plots are given in Figure 2. We conclude that the non-supersymmetric $AdS_7$ fixed point is $not$ stable.

\begin{figure}[t]
\begin{center}
\includegraphics[width=3.0in]{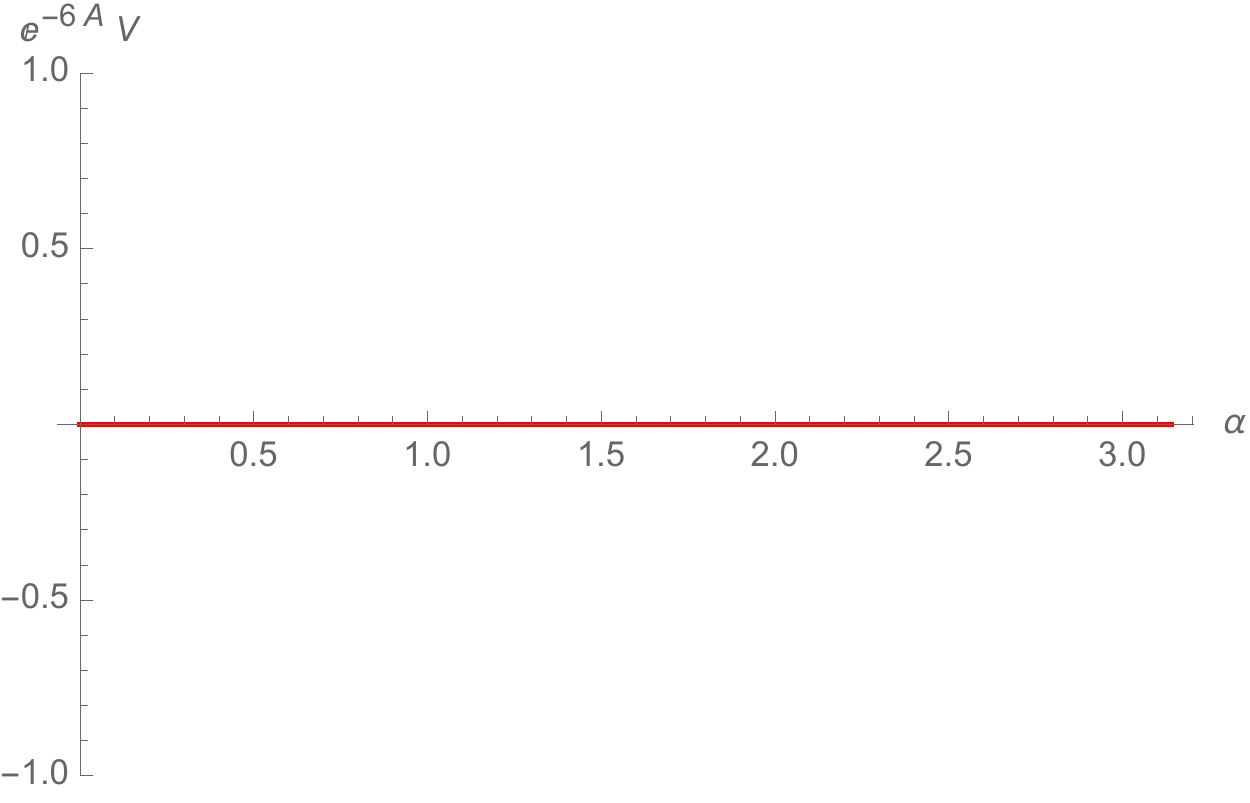} \qquad \includegraphics[width=3.0in]{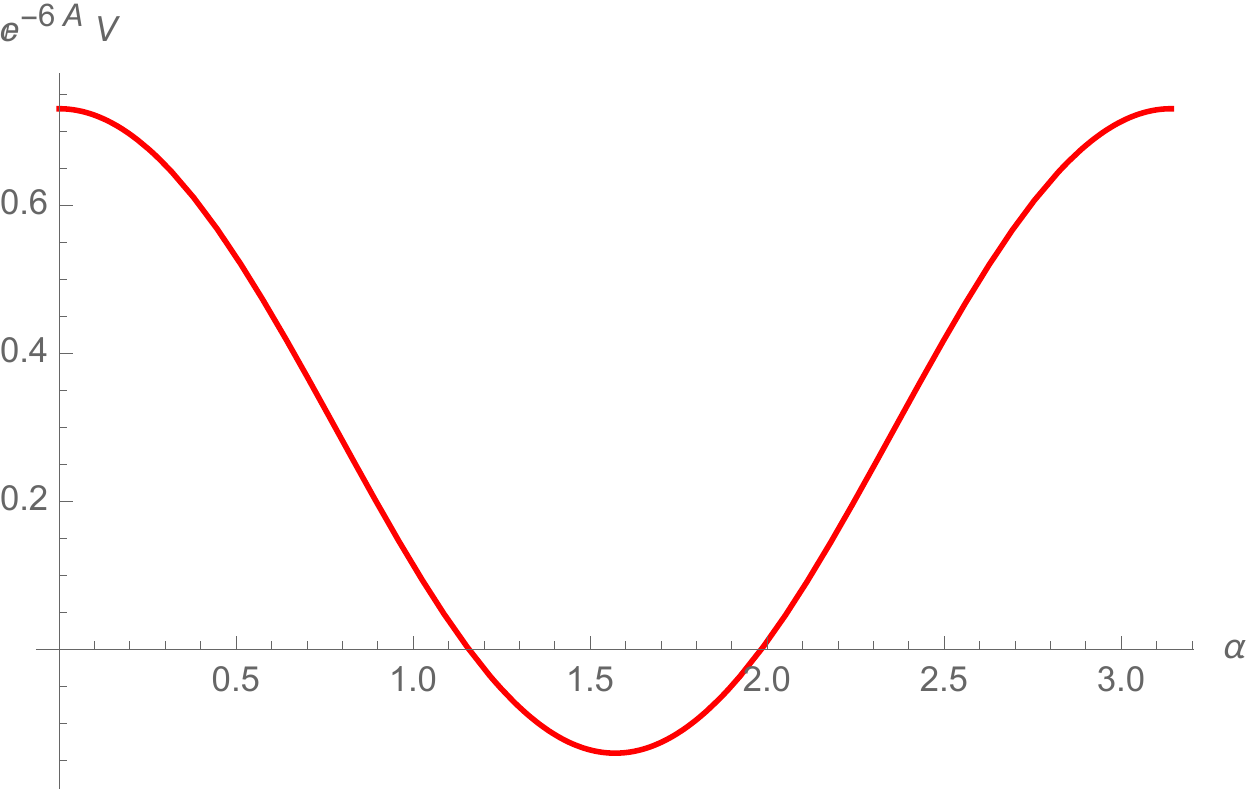}
\caption{{\it The probe brane potentials of the supersymmetric and non-supersymmetric fixed points at $X\,=\,1$ and $X\,=\,1/2^{1/5}$, respectively.}}
\label{1}
\end{center}
\end{figure}

\bigskip
\bigskip
\leftline{\bf Acknowledgements}
\noindent We are grateful to Nakwoo Kim for helpful discussions and to Krzysztof Pilch for reading a draft. We would like to thank Gabriele Lo Monaco and collaborators of \cite{Apruzzi:2019ecr} for comments on a preprint. This research was supported by the National Research Foundation of Korea under the grant  NRF-2019R1I1A1A01060811.

\vspace{3.4cm}

\appendix
\section{Potentials of the fluxes for supersymmetric flows}
\renewcommand{\theequation}{A.\arabic{equation}}
\setcounter{equation}{0} 

For the flows to supersymmetric fixed points, we can derive the potentials of the fluxes not just at the fixed point but over the whole flow. In the appendix we present the derivations.

\subsection{Flows from $AdS_6$}

We consider the domain wall background, \cite{Gursoy:2002tx},
\begin{equation}
ds_6^2\,=\,e^{2A}ds_{1,4}^2+dr^2\,.
\end{equation}
The supersymmetry equations are given by
\begin{align} \label{susyeq}
\phi'\,=&\,g\left(e^{-\frac{\phi}{2\sqrt{2}}}-e^{\frac{3\phi}{2\sqrt{2}}}\right)\,, \notag \\
A'\,=&\,\frac{g}{2\sqrt{2}}\left(e^{-\frac{\phi}{2\sqrt{2}}}+\frac{1}{3}e^{\frac{3\phi}{2\sqrt{2}}}\right)\,.
\end{align}

The uplift formula for the six-form flux is given by, \cite{Cvetic:1999un},
\begin{equation}
F_{(6)}\,=\,e^{\Phi/2}*F_{(4)}\,=\,-\frac{\sqrt{2}g}{3}Uvol_6+\frac{4\sqrt{2}}{g}\frac{\sin\xi\cos\xi}{X}*dX\wedge{d}\xi\,.
\end{equation}
For the domain wall solutions, the six-form flux is
\begin{equation}
F_{(6)}\,=\,\omega_r\,dx_0\wedge{d}x_1\wedge{d}x_2\wedge{d}x_3\wedge{d}x_4\wedge{d}r+\omega_\xi\,dx_0\wedge{d}x_1\wedge{d}x_2\wedge{d}x_3\wedge{d}x_4\wedge{d}\xi\,,
\end{equation}
where
\begin{align}
\omega_r\,=&\,-\frac{\sqrt{2}g}{3}e^{5A}U\,, \notag \\
\omega_\xi\,=&\,\frac{4\sqrt{2}}{g}\frac{e^{5A}X'\sin\xi\cos\xi}{X}\,.
\end{align}
Employing the supersymmetry equations, \eqref{susyeq}, they satisfy a relation,{\footnote{This is an analogous calculation of (3.13), (3.14), (3.28), (3.29) from \cite{Pilch:2000fu} and (8) from \cite{Johnson:2000ic}.}}
\begin{equation}
\frac{\partial\omega_\xi}{\partial{r}}\,=\,\frac{\partial\omega_r}{\partial\xi}\,.
\end{equation}
Then we obtain that the five-form potential is
\begin{equation}
C_{(5)}\,=\,-e^{5A}\Delta\,dx_0\wedge{d}x_1\wedge{d}x_2\wedge{d}x_3\wedge{d}x_4\,.
\end{equation}
If we employ this five-form potential to compute the probe brane potential, it vanishes identically over the whole flow,
\begin{equation}
V\,=\,e^{5A}\left(\Delta-\Delta\right)\,=\,0\,.
\end{equation}

\subsection{Flows from $AdS_7$}

We consider the domain wall background, \cite{Campos:2000yu},
\begin{equation}
ds_7^2\,=\,e^{2A}ds_{1,5}^2+dr^2\,.
\end{equation}
The supersymmetry equations are given by
\begin{align} \label{susyeq7}
\lambda'\,=&\,\frac{2}{5}e^{-8\lambda}-\frac{2}{5}e^{2\lambda}\,, \notag \\
A'\,=&\,\frac{1}{5}e^{-8\lambda}+\frac{4}{5}e^{2\lambda}\,.
\end{align}

The uplift formula for the seven-form flux is given by, \cite{Cvetic:1999xp},
\begin{equation}
F_{(7)}\,=\,Uvol_7+\frac{1}{2g}\sum_{\alpha=0}^2X_\alpha^{-1}*_7dX_\alpha\wedge{d}(\mu_\alpha^2)\,,
\end{equation}
From an analogous calculation of the previous subsection, we obtain that the six-form potential is
\begin{equation}
C_{(6)}\,=\,-e^{6A}\Delta\,dx_0\wedge{d}x_1\wedge{d}x_2\wedge{d}x_3\wedge{d}x_4\wedge{d}x_5\,.
\end{equation}
If we employ this six-form potential to compute the probe brane potential, it vanishes identically over the whole flow,
\begin{equation}
V\,=\,e^{6A}\left(\Delta-\Delta\right)\,=\,0\,.
\end{equation}




\begin{thebibliography}{99}

\bibitem{Maldacena:1997re}
  J.~M.~Maldacena,
  {\it The large N limit of superconformal field theories and supergravity,}
  Adv.\ Theor.\ Math.\ Phys.\  {\bf 2}, 231 (1998)
  [Int.\ J.\ Theor.\ Phys.\  {\bf 38}, 1113 (1999)]
  [arXiv:hep-th/9711200].

\bibitem{Breitenlohner:1982bm} 
  P.~Breitenlohner and D.~Z.~Freedman,
  {\it Positive Energy in anti-De Sitter Backgrounds and Gauged Extended Supergravity,}
  Phys.\ Lett.\  {\bf 115B}, 197 (1982).

\bibitem{Breitenlohner:1982jf} 
  P.~Breitenlohner and D.~Z.~Freedman,
  {\it Stability in Gauged Extended Supergravity,}
  Annals Phys.\  {\bf 144}, 249 (1982).

\bibitem{Gibbons:1983aq} 
  G.~W.~Gibbons, C.~M.~Hull and N.~P.~Warner,
  {\it The Stability of Gauged Supergravity,}
  Nucl.\ Phys.\ B {\bf 218}, 173 (1983).

\bibitem{BenettiGenolini:2019zth} 
  P.~Benetti Genolini, M.~Honda, H.~C.~Kim, D.~Tong and C.~Vafa,
  {\it Evidence for a Non-Supersymmetric 5d CFT from Deformations of 5d $SU(2)$ SYM,}
  arXiv:2001.00023 [hep-th].

\bibitem{ArkaniHamed:2006dz} 
  N.~Arkani-Hamed, L.~Motl, A.~Nicolis and C.~Vafa,
  {\it The String landscape, black holes and gravity as the weakest force,}
  JHEP {\bf 0706}, 060 (2007) [hep-th/0601001].

\bibitem{Ooguri:2016pdq} 
  H.~Ooguri and C.~Vafa,
  {\it Non-supersymmetric AdS and the Swampland,}
  Adv.\ Theor.\ Math.\ Phys.\  {\bf 21}, 1787 (2017) [arXiv:1610.01533 [hep-th]].

\bibitem{Bena:2020xxb} 
  I.~Bena, K.~Pilch and N.~P.~Warner,
  {\it Brane-Jet Instabilities,}
  arXiv:2003.02851 [hep-th].

\bibitem{Warner:1983du} 
  N.~P.~Warner,
  {\it Some Properties of the Scalar Potential in Gauged Supergravity Theories,}
  Nucl.\ Phys.\ B {\bf 231}, 250 (1984).

\bibitem{Warner:1983vz} 
  N.~P.~Warner,
  {\it Some New Extrema of the Scalar Potential of Gauged $N=8$ Supergravity,}
  Phys.\ Lett.\  {\bf 128B}, 169 (1983).

\bibitem{deWit:1982bul} 
  B.~de Wit and H.~Nicolai,
  {\it N=8 Supergravity,}
  Nucl.\ Phys.\ B {\bf 208}, 323 (1982).

\bibitem{Khavaev:1998fb} 
  A.~Khavaev, K.~Pilch and N.~P.~Warner,
  {\it New vacua of gauged N=8 supergravity in five-dimensions,}
  Phys.\ Lett.\ B {\bf 487}, 14 (2000) [hep-th/9812035].

\bibitem{Guarino:2020jwv} 
  A.~Guarino, J.~Tarrio and O.~Varela,
  {\it Brane-jet stability of non-supersymmetric AdS vacua,}
  arXiv:2005.07072 [hep-th].

\bibitem{Maldacena:1998uz}
J.~M.~Maldacena, J.~Michelson and A.~Strominger,
{\it Anti-de Sitter fragmentation,}
JHEP \textbf{02}, 011 (1999) [arXiv:hep-th/9812073 [hep-th]].

\bibitem{Apruzzi:2019ecr} 
  F.~Apruzzi, G.~Bruno De Luca, A.~Gnecchi, G.~Lo Monaco and A.~Tomasiello,
  {\it On AdS$_7$ stability,}
  arXiv:1912.13491 [hep-th].

\bibitem{Horowitz:2007pr}
G.~T.~Horowitz, J.~Orgera and J.~Polchinski,
{\it Nonperturbative Instability of AdS(5) x S**5/Z(k),}
Phys. Rev. D \textbf{77}, 024004 (2008) [arXiv:0709.4262 [hep-th]].

\bibitem{Gaiotto:2009mv}
D.~Gaiotto and A.~Tomasiello,
{\it The gauge dual of Romans mass,}
JHEP \textbf{01}, 015 (2010) [arXiv:0901.0969 [hep-th]].

\bibitem{Narayan:2010em}
P.~Narayan and S.~P.~Trivedi,
{\it On The Stability Of Non-Supersymmetric AdS Vacua,}
JHEP \textbf{07}, 089 (2010) [arXiv:1002.4498 [hep-th]].

\bibitem{Ooguri:2017njy}
H.~Ooguri and L.~Spodyneiko,
{\it New Kaluza-Klein instantons and the decay of AdS vacua,}
Phys. Rev. D \textbf{96}, no.2, 026016 (2017) [arXiv:1703.03105 [hep-th]].

\bibitem{Romans:1985tw} 
  L.~J.~Romans,
  {\it The F(4) Gauged Supergravity in Six-dimensions,}
  Nucl.\ Phys.\ B {\bf 269}, 691 (1986).

\bibitem{Townsend:1983kk} 
  P.~K.~Townsend and P.~van Nieuwenhuizen,
  {\it Gauged Seven-dimensional Supergravity,}
  Phys.\ Lett.\  {\bf 125B}, 41 (1983).

\bibitem{Mezincescu:1984ta} 
  L.~Mezincescu, P.~K.~Townsend and P.~van Nieuwenhuizen,
  {\it Stability of Gauged $d=7$ Supergravity and the Definition of Masslessness in ({AdS}) in Seven-dimensions,}
  Phys.\ Lett.\  {\bf 143B}, 384 (1984).

\bibitem{Pernici:1984xx} 
  M.~Pernici, K.~Pilch and P.~van Nieuwenhuizen,
  {\it Gauged Maximally Extended Supergravity in Seven-dimensions,}
  Phys.\ Lett.\  {\bf 143B}, 103 (1984).

\bibitem{Pernici:1984zw} 
  M.~Pernici, K.~Pilch, P.~van Nieuwenhuizen and N.~P.~Warner,
  {\it Noncompact Gaugings and Critical Points of Maximal Supergravity in Seven-dimensions,}
  Nucl.\ Phys.\ B {\bf 249}, 381 (1985).

\bibitem{Nastase:1999cb} 
  H.~Nastase, D.~Vaman and P.~van Nieuwenhuizen,
  {\it Consistent nonlinear K K reduction of 11-d supergravity on AdS(7) x S(4) and selfduality in odd dimensions,}
  Phys.\ Lett.\ B {\bf 469}, 96 (1999) [hep-th/9905075].

\bibitem{Nastase:1999kf} 
  H.~Nastase, D.~Vaman and P.~van Nieuwenhuizen,
  {\it Consistency of the AdS(7) x S(4) reduction and the origin of selfduality in odd dimensions,}
  Nucl.\ Phys.\ B {\bf 581}, 179 (2000)[hep-th/9911238].

\bibitem{Nastase:2000tu} 
  H.~Nastase and D.~Vaman,
  {\it On the nonlinear KK reductions on spheres of supergravity theories,}
  Nucl.\ Phys.\ B {\bf 583}, 211 (2000) [hep-th/0002028].

\bibitem{Lu:1999bc} 
  H.~Lu and C.~N.~Pope,
  {\it Exact embedding of N=1, D = 7 gauged supergravity in D = 11,}
  Phys.\ Lett.\ B {\bf 467}, 67 (1999) [hep-th/9906168].

\bibitem{Passias:2015gya} 
  A.~Passias, A.~Rota and A.~Tomasiello,
  {\it Universal consistent truncation for 6d/7d gauge/gravity duals,}
  JHEP {\bf 1510}, 187 (2015) [arXiv:1506.05462 [hep-th]].

\bibitem{Cvetic:1999un} 
  M.~Cvetic, H.~Lu and C.~N.~Pope,
  {\it Gauged six-dimensional supergravity from massive type IIA,}
  Phys.\ Rev.\ Lett.\  {\bf 83}, 5226 (1999) [hep-th/9906221].

\bibitem{Jeong:2013jfc} 
  J.~Jeong, O.~Kelekci and E.~O Colgain,
  {\it An alternative IIB embedding of F(4) gauged supergravity,}
  JHEP {\bf 1305}, 079 (2013) [arXiv:1302.2105 [hep-th]].

\bibitem{Hong:2018amk} 
  J.~Hong, J.~T.~Liu and D.~R.~Mayerson,
  {\it Gauged Six-Dimensional Supergravity from Warped IIB Reductions,}
  JHEP {\bf 1809}, 140 (2018) [arXiv:1808.04301 [hep-th]].

\bibitem{Malek:2018zcz} 
  E.~Malek, H.~Samtleben and V.~Vall Camell,
  {\it Supersymmetric AdS$_{7}$ and AdS$_6$ vacua and their minimal consistent truncations from exceptional field theory,}
  Phys.\ Lett.\ B {\bf 786}, 171 (2018) [arXiv:1808.05597 [hep-th]].

\bibitem{DHoker:2016ujz}
E.~D'Hoker, M.~Gutperle, A.~Karch and C.~F.~Uhlemann,
{\it Warped $AdS_6\times S^2$ in Type IIB supergravity I: Local solutions,}
JHEP \textbf{08} (2016), 046 [arXiv:1606.01254 [hep-th]].

\bibitem{DHoker:2016ysh}
E.~D'Hoker, M.~Gutperle and C.~F.~Uhlemann,
{\it Holographic duals for five-dimensional superconformal quantum field theories,}
Phys. Rev. Lett. \textbf{118} (2017) no.10, 101601 [arXiv:1611.09411 [hep-th]].

\bibitem{Apruzzi:2013yva}
F.~Apruzzi, M.~Fazzi, D.~Rosa and A.~Tomasiello,
{\it All AdS$_7$ solutions of type II supergravity,}
JHEP {\bf 04} (2014), 064 [arXiv:1309.2949 [hep-th]].

\bibitem{Apruzzi:2016rny} 
  F.~Apruzzi, G.~Dibitetto and L.~Tizzano,
  {\it A new 6d fixed point from holography,}
  JHEP {\bf 1611}, 126 (2016) [arXiv:1603.06576 [hep-th]].

\bibitem{Danielsson:2017max} 
  U.~H.~Danielsson, G.~Dibitetto and S.~C.~Vargas,
  {\it A swamp of non-SUSY vacua,}
  JHEP {\bf 1711}, 152 (2017) [arXiv:1708.03293 [hep-th]].

\bibitem{Romans:1985tz} 
  L.~J.~Romans,
  {\it Massive N=2a Supergravity in Ten-Dimensions,}
  Phys.\ Lett.\ B {\bf 169}, 374 (1986).

\bibitem{Brandhuber:1999np} 
  A.~Brandhuber and Y.~Oz,
  {\it The D-4 - D-8 brane system and five-dimensional fixed points,}  Phys.\ Lett.\ B {\bf 460}, 307 (1999) [hep-th/9905148].

\bibitem{Cvetic:1999xp} 
  M.~Cvetic, M.J.~Duff, P.~Hoxha, J.T.~Liu, H.~Lu, J.X.~Lu, R.~Martinez-Acosta, C.N.~Pope, H.~Sati, and T.A.~Tran,
  {\it Embedding AdS black holes in ten-dimensions and eleven-dimensions,}
  Nucl.\ Phys.\ B {\bf 558}, 96 (1999) [hep-th/9903214].

\bibitem{Cremmer:1978km} 
  E.~Cremmer, B.~Julia and J.~Scherk,
  {\it Supergravity Theory in Eleven-Dimensions,}
  Phys.\ Lett.\  {\bf 76B}, 409 (1978).

\bibitem{Gursoy:2002tx} 
  U.~Gursoy, C.~Nunez and M.~Schvellinger,
  {\it RG flows from spin(7), CY 4 fold and HK manifolds to AdS, Penrose limits and pp waves,}
  JHEP {\bf 0206}, 015 (2002) [hep-th/0203124].

\bibitem{Pilch:2000fu} 
  K.~Pilch and N.~P.~Warner,
  {\it N=1 supersymmetric renormalization group flows from IIB supergravity,}
  Adv.\ Theor.\ Math.\ Phys.\  {\bf 4}, 627 (2002) [hep-th/0006066].

\bibitem{Johnson:2000ic} 
  C.~V.~Johnson, K.~J.~Lovis and D.~C.~Page,
  {\it Probing some N=1 AdS / CFT RG flows,}
  JHEP {\bf 0105}, 036 (2001) [hep-th/0011166].

\bibitem{Campos:2000yu} 
  V.~L.~Campos, G.~Ferretti, H.~Larsson, D.~Martelli and B.~E.~W.~Nilsson,
  {\it A Study of holographic renormalization group flows in D = 6 and D = 3,}
  JHEP {\bf 0006}, 023 (2000) [hep-th/0003151].

\end{thebibliography}
\end{document}